\documentclass[fleqn,usenatbib]{mnras}

\usepackage{newtxtext,newtxmath}

\usepackage[T1]{fontenc}
\usepackage{ae,aecompl}

\usepackage{graphicx}	
\usepackage{amsmath}	
\usepackage{amssymb}	

\usepackage{epstopdf}
\usepackage[pdftex]{color}
\usepackage{colortbl}
\usepackage[table]{xcolor}

\title[A wavelet analysis of white dwarf stars]{A wavelet analysis  of photometric variability in {\em Kepler} white dwarf stars}

\author[S. R. de Lira et al.]{
S. R. de Lira,$^{1}$\thanks{E-mail: suzierly@ufrn.edu.br}
J. P. Bravo,$^{1,2}$
I. C. Le\~ao,$^{1}$ 
A. D. da Costa, $^{1,3}$
B. L. Canto Martins,$^{1}$\newauthor 
D. B. de Freitas$^{5}$ and
J. R. De Medeiros$^{1}$
\\
$^{1}$Universidade Federal do Rio Grande do Norte, Departamento de F\'{\i}sica Te\'orica e Experimental, 59072-970 Natal, RN, Brazil\\
$^{2}$Instituto Federal de Educa\c{c}\~ao, Ci\^encia e Tecnologia do Rio Grande do Norte, Natal, RN 59015-000, Brazil\\
$^{3}$Universidade da Integra\c{c}\~ao Internacional da Lusofonia Afro-Brasileira, Reden\c{c}\~ao, CE 62790-000, Brazil\\
$^{5}$Universidade Federal do Cear\'a, Departamento de F\'{\i}sica, Campus do Pici, Fortaleza, CE 60455-900, Brazil\\
}

\date{Accepted XXX. Received YYY; in original form ZZZ}

\pubyear{2019}

\begin{document}
\label{firstpage}
\pagerange{\pageref{firstpage}--\pageref{lastpage}}
\maketitle

\begin{abstract}

This work brings a wavelet analysis for 14 {\it Kepler} white dwarf stars, in order to confirm their photometric variability behavior and to search for periodicities in these targets. From the observed {\it Kepler} light curves we obtained the wavelet local and global power spectra. Through this procedure, one can perform an analysis in time-frequency domain rich in details, and so to obtain a new perspective on the time evolution of the periodicities present in these stars.  We identified a photometric variability behavior in ten white dwarfs, corresponding to period variations of $\sim$ 2 h to 18 days: among these stars, three are new candidates and seven, earlier identified from other studies, are confirmed.

\end{abstract}

\begin{keywords}
stars: white dwarfs -- methods: data analysis -- techniques: photometric
\end{keywords}



\section{Introduction}

\begin{table*}
\begin{minipage}{\textwidth}
\centering
\caption{White dwarfs parameters, including photometric periodic variations computed via wavelet technique.}
\label{t_data}
\def\arraystretch{1.18} 
\begin{tabular}{|p{1.5cm}ccccc|cc|}
\hline 
KIC &  Spectral & $\mathrm{K_p}$ & $\mathrm{T_{eff}}$ & $\mathrm{\log g}$ & Period \footnote{Periods obtained by \citet{Maoz2015}}& Quarters\footnote{Quarters of long cadence available in \emph{Kepler} public data and analyzed in this work} & Wavelet Variability Periods\\  & type &(mag)& (kK) & (cm $\mathrm{s^{-2}}$) &  (days) &  & (days)\\ 
\hline
\multicolumn{6}{c}{Periodic \citep{Maoz2015}}&\multicolumn{2}{c}{Our analysis}\\
\hline
5769827 &  DA & 16.6 & 66 & 8.2  & 8.30$\pm0.04$ & 4,6 &  8.02\\
6669882 &  DA & 17.9 & 30.5 &	7.4 & 0.367$\pm$0.009 & 2 &  0.36\\
6862653 &  subdwarf & 18.2 & 40.3 & 6.4 & 0.594$\pm$0.02 & 2 & 0.60\\
8682822 &  DA &15.8 & 23.1 & 8.5 & 4.7$\pm$0.3 & 5-9 & 5.29\\
11337598 & DA & 16.1 & 22.8 & 8.6 & 0.09328$\pm$0.00003 & 3,10 & 0.09\\
11514682 & DA & 15.7 & 32.2 & 7.5 & 9.89$\pm$0.06 & 2-17 & 17.59\\
11604781 & DA  & 16.7 & 9.1 & 8.3 & 4.89$\pm$0.02 & 3,6,7,10,11 & 4.73\\
\hline
\multicolumn{6}{c}{Non-periodic \citep{Maoz2015}}&\multicolumn{2}{c}{Our analysis} \\
\hline
3427482 &  DA & 17.3 & - & - & - & 1 & -\\
4829241 & DA & 15.8 & 19.5 & 8.0 & - & 2-13 & 16.04 \\
7129927 & composite DA+DA & 16.6 & 9.5 & 8.3 & - & 3,5,6 & -\\
9139775 &  DA & 17.9 & 24.6 & 8.6 & - & 2 &  -\\
10198116 &  DA & 16.4 & 13.5 & 8.0 & - & 4-6 &  -\\ 
10420021 &  DA & 16.2 &  12.8 & 7.8 & - & 2,5,6,8-10 & 11.70\\
11822535 &  DA & 14.8 & 36.0 & 7.9 & - & 2-13 & 13.83\\
\hline
\end{tabular}
\end{minipage}
\end{table*}

The observations of the space missions {\it Kepler} and {\it CoRoT} are revolutionizing our understanding of stellar variability and generating new insights in the correlation between photometric variability and rotation of main sequence stars \citep[e.g.][]{Davenport2017,Leao2015,Chinchon2015,McQuillan2014,DeMedeiros2013,Nielsen2013,W2013} and beyond the main sequence \citep[e.g.][]{Costa2015,Saders2013,DeMedeiros2013,Mosser2012}. These works have shown that rotation is a major constraint on the study of the net angular momentum evolution as well as on the angular momentum transport from core to surface and expanding envelope. In addition, the normalcy of Sun rotation with respect to the main sequence stars with surface physical parameters close to the solar values \citep{Leao2015,deFreitas2013} and bimodality in the rotation period distribution for main sequence stars \citep{Davenport2017,McQuillan2014} have also emerged from data acquired by the referred space missions. 

The observation of photometric modulation has also a unique potential to probe for rotation in white dwarfs (hereafter WD), one of the few remaining traces to the physics of the formation process of WD \citep{Kawaler2015,Kawaler2003}. Such a procedure is more easily applied to magnetic white dwarfs, which comprise approximately 20 percent of white dwarf stars presently reported in the literature \citep[e.g.][]{Kepler2013,Kawka2007}. Typically, the observed rotation period for these stars ranges from less than one hour to a few days \citep[e.g.][]{Kawaler2015,Brinkworth2013,Kawka2007}. Nevertheless, photometric modulation in normal white dwarfs can also be revealed by surface dynamo activity, producing localized star-spots \citep[e.g.][]{Kawaler2015}. Finally, asteroseismology analysis, based on high photometric precision, can be used for the measurements of the rotation period of white dwarfs, showing the appropriate range of effective temperature in which they undergo non-radial g-mode pulsations. 

\citet{Maoz2015} have conducted a Fast Fourier Transform analysis of Kepler light curves (hereafter LCs)  in 14 WD, searching for photometric variability. Those authors detected periodic signals ranging from 2\thinspace hours to 10\thinspace days for 7 among the 14 targets they analyzed. These signals were detected in both short and long cadence data LCs from \emph{Kepler} Mission.  In the present work, we perform a time-frequency analysis of the same Kepler WD datasets
studied by \citet{Maoz2015} in the interest to point out the spectral components, and to follow their time evolution which leads to short- and long-lived signatures formation that could be linked to the WD rotation or activity. For this purpose, since the continuous wavelet transform (WT) is a powerful mathematical tool for non-stationary signals, we used a 6th order Morlet wavelet \citep{Torrence1998} to obtain the energy distribution of each LC computed as a local wavelet spectrum, mapping short- and long-term features.

\vspace{0.1cm}

This paper is organized as follows. Section \ref{sample} presents our stellar sample and the methods used for this analysis. Section \ref{results} brings the primary results, in which  the periods detected from the wavelet method are analysed and compared with those obtained by \citet{Maoz2015}. Finally, our conclusions are provided in Section \ref{conclusions}.

\section{Sample selection and LC analysis}\label{sample}

NASA \emph{Kepler} mission provides LCs in 18 runs (also known as \emph{quarters}) with standard treatment as Simple Aperture Photometry (SAP) data and with a more refined treatment, in which instrumental effects were removed, as Pre-search Data Conditioning (PDC) \citep{Stumpe2012,Smith2012}. For our purposes, we have chosen 14 WD with long (30 minutes) and short (1 minute) cadence LCs retrieved from the \emph{Kepler} Data Archive\footnote{http://archive.stsci.edu/kepler/}. Those targets were previously studied by \citet{Ostensen2011,Ostensen2010} searching for pulsation among compact objects, and by \citet{Maoz2015} focusing on photometric variations using the FFT.

As underlined by \citet{Maoz2015}, their analysis made use of Public Releases $\mathrm{14 - 21}$ of the long cadence data and Public Release 21 of the short cadence, downloaded from the {\em Kepler} mission archive. However, it seems that no additional treatment of the LCs was performed by the authors. The detection threshold for periodicity varies from star-to-star, depending on its brightness, the number of quarters of observations, and the final cleaning of the LCs. To avoid possible distortions in the spectral wavelet maps, and consequently, changes in the power spectra, an additional treatment as outliers removal and instrumental trend corrections was performed in the PDC LCs \citep[e.g.][]{Chinchon2015}. The single long-term LC was obtained by assembling the available quarters for each object, based on the procedure of \citet{Banyai2013}. The map signatures are then well defined, inducing to more reliable variability periodicities.

Wavelet power spectra were analysed based on procedures described by \citet{Bravo2014}. The local map is defined as the signal energy distribution in time and frequency, while the global power spectrum is computed from the time integration of the wavelet map. Examples of these are displayed in the Results section. The WT application and the variability analysis was performed for each quarter of the whole sample, as well as for merged LCs in short- and long-cadence modes. Since WD time series present some gaps, the wavelet map was obtained individually for each set of continuous data. Thus, the global spectrum was produced by considering the weighted average by time span of these parts \citep{Costa2015} for the targets KIC 4829241, KIC 5769827, KIC 6669882, KIC 6862653, KIC 8682822, KIC 10420021, KIC 11604781 and KIC 11822535. For the referred stars, the data treatment is related to long cadence data set. The objects of our sample are listed in Table \ref{t_data}, including their {\it Kepler} magnitude, $K_p$; theirtheir spectral type, effective temperature, $T_{eff}$, and surface gravity, $\log g$, \citep[e.g.][]{2018MNRAS.476..933H,2017MNRAS.464.3464D,Ostensen2011}; the periodicities computed by \citet{Maoz2015} using the FFT and those obtained using our wavelet procedure.
     
Furthermore, we have applied the FFT in our data after additional treatments to confirm the periods reported in Table \ref{t_data}, as well as the Lomb-Scargle method \citep{1982ApJ...263..835S} to verify the variability nature of some targets which seems to be quasi-periodic after a visual inspection of the wavelet map signature although classified as non-periodic by \citet{Maoz2015}. In each Lomb-Scargle periodogram, main peaks were identified with a false alarm probability (FAP) \citep{Horne1986} less than 0.01, or a significance level greater than 99\%.
This FAP estimation was used with caution just as a qualitative indication of reliability because it assumes a random noise in the LCs, which actually have more complex noise signatures.

\begin{figure}
	\includegraphics[width=\columnwidth]{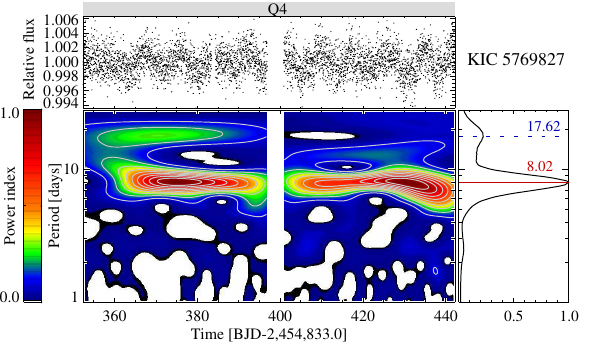}
	\includegraphics[width=\columnwidth]{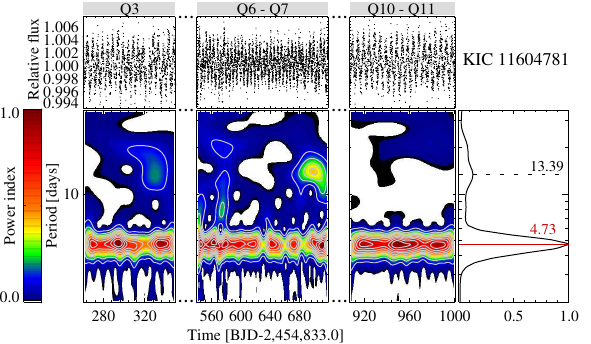}
    \caption{Light curves with additional treatment, wavelet local and global spectra of KIC 5769827 (upper panel) and KIC 11604781 (lower panel). Contour levels are 90$\%$, 80$\%$, ..., 20$\%$ and 10$\%$ of the map maximum. The variability periods are illustrated by the red dashed lines in the global spectra. Plausible aliasing period is displayed in blue color.}
    \label{fig1}
\end{figure}

\begin{figure}
	\includegraphics[width=\columnwidth]{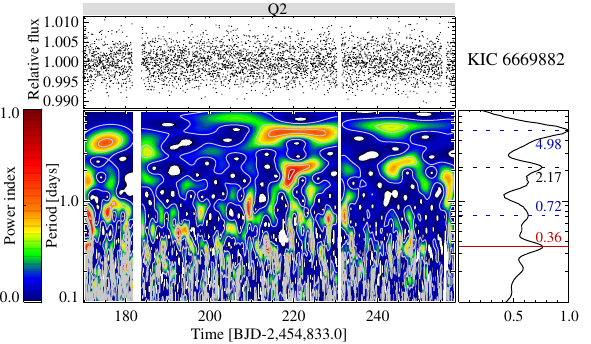}
	\includegraphics[width=\columnwidth]{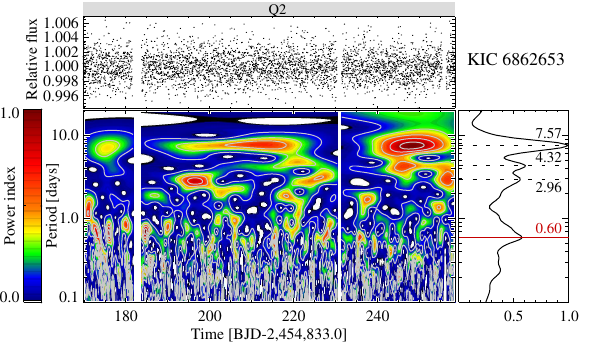}
	\includegraphics[width=\columnwidth]{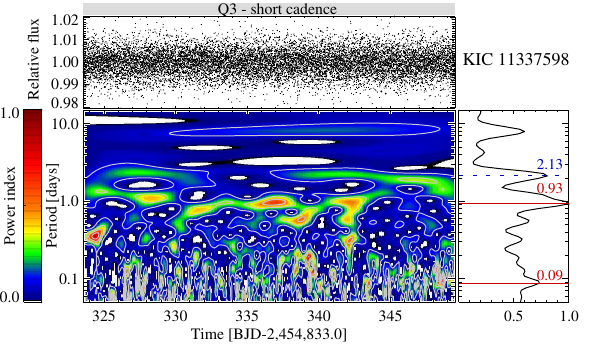}
    \caption{Light curves with additional treatment, wavelet local and global spectra of {\em Kepler} WD which were classified as periodic by \citet{Maoz2015}, and their results were reproduced via wavelet (from top to bottom) for KIC 6669882, KIC 6862653, and KIC 11337598. Contour levels are 90$\%$, 80$\%$, ..., 20$\%$ and 10$\%$ of the map maximum. The variability periods are illustrated by the red dashed lines in the global spectra. Plausible aliasing period is displayed in blue color.}
    \label{fig2}
\end{figure}

\section{Results and discussion}\label{results}

\begin{figure}
	\includegraphics[width=\columnwidth]{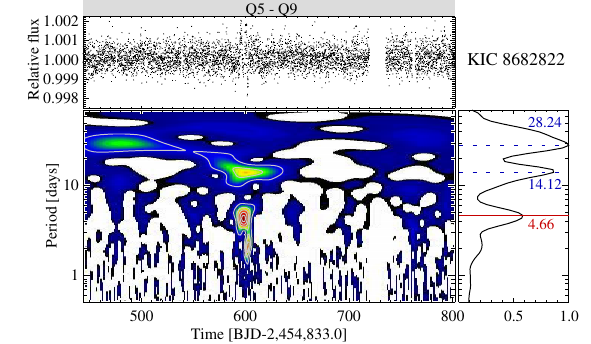}
	\includegraphics[width=\columnwidth]{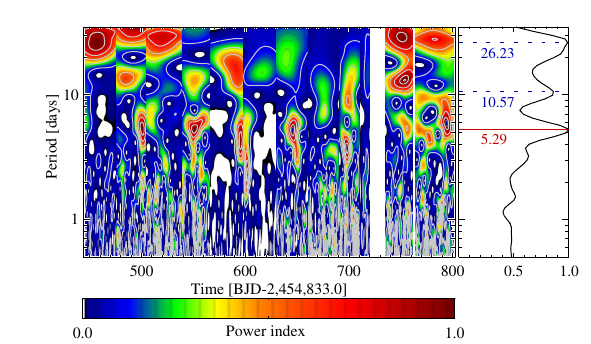}
    \caption{{\it Upper panel}: Light curve with additional treatment, wavelet local and global spectra of KIC 8682822. The period of 4.66\thinspace days with low amplitude, illustrated by a red solid line, is present in a short time interval, which is not enough to confirm a periodic variability. {\it Lower panel}: wavelet local spectra of KIC 8682822 of different short datasets. After assemblage, the weighted average by time span points to a 5-day periodicity with highest amplitude (red solid line). Contour levels are 90$\%$, 70$\%$, 50$\%$, 30$\%$, and 10$\%$ of the map maximum. The 11-, 14-, 28-, and 26-day periodicities, displayed in blue color in both maps, are possibly aliasing periods.}

\label{fig3}
\end{figure}

Considering that the observed rotation periods in white dwarfs range from less than one hour to a few days \citep{Kawaler2015}, in the present work we are mostly searching for short periodicities.
Our wavelet procedure confirms the periodic modulation found by \citet{Maoz2015} for the stars KIC 5769827, KIC 6669882, KIC 6862653, KIC 8682822, KIC 11337598 and KIC 11604781, with periodicities similar or close to the values reported by those authors, as illustrated by their wavelet maps in Figures \ref{fig1} to \ref{fig3}. Nevertheless, for the WD KIC 11514682 we found a period of 17.59 days, in contrast with \citet{Maoz2015} who computed a 9.89\thinspace days period. Among the seven WD stars classified by \citet{Maoz2015} as non-periodic, our wavelet analysis reveals unambiguous periodicities for three of them, KIC 4829241, KIC 10420021 and KIC 11822535, with 16.04\thinspace days, 11.70\thinspace days and 13.83\thinspace days, respectively. The computed wavelet periodicities, along with those from \citet{Maoz2015}, are listed in Table \ref{t_data}. 

The results obtained using the long-cadence data are in agreement with those found making use of short-cadence data. However, for better viewing of the variability signature in the WD wavelet maps, and hence a clear identification of periodicities, we given priority to display long-cadence datasets. Thus, the wavelet periods in Table \ref{t_data}, also displayed in the respective wavelet global spectra, are related to the Kepler long-cadence mode. One exception is the WD KIC 11337598, for which variability is better defined in the wavelet spectra using short-cadence data.

Let us underline that, except for KIC 8682822 and KIC 11822535, the level of persistence of the periodicity in the wavelet maps of the analysed stars is larger than 50\%, a fairly significant fraction of the entire data set. For KIC 8682822 and KIC 11822535 the fraction persistence is about 5\% and 40\%, respectively. In the following we discuss a few particular features emerging from our wavelet analysis, in particular for those stars with photometric variability revealed in the present study.

\subsection{KIC 11337598} 

\citet{Ostensen2011} detected a Balmer line broadening in the spectrum of KIC 11337598, suggesting the possibility that this WD is a rapid rotator with a $v \sin i=1500$\,km s$^{-1}$. The fit used by those authors corresponds to a rotational period of about 
$\mathrm{40\,s}$ or $\mathrm{0.000463\,days}$.  Moreover, \citet{Maoz2015} suggested a period of 0.09\thinspace days as the variability period for KIC 11337598. In Figure \ref{fig2} (lower panel), our wavelet analysis  reveals three dominant periods,  $\mathrm{P_1=0.93}$\thinspace days, $\mathrm{P_2=2.13}$\thinspace days and $\mathrm{P_3=0.09}$\thinspace days. The period $\mathrm{P_3}$ is persistent over the entire time span, corresponding to the main wavelet period, in agreement with \citet{Maoz2015}. $\mathrm{P_1}$ is also persistent, and probably it is a P3 multiple period related to rotation. It should be noted that $\mathrm{P_1}$ is exactly 2000 times the 40-second periodicity found by \citet{Ostensen2011}. Therefore, we consider here both $\mathrm{P_1}$ and $\mathrm{P_3}$ linked to the WD rotation, while $\mathrm{P_2}$ is a plausible aliasing period.
 
\begin{figure}
	\includegraphics[width=\columnwidth]{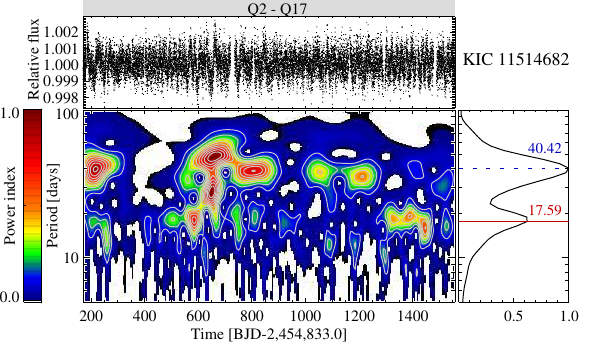}
	\includegraphics[height=4.75cm,width=\columnwidth]{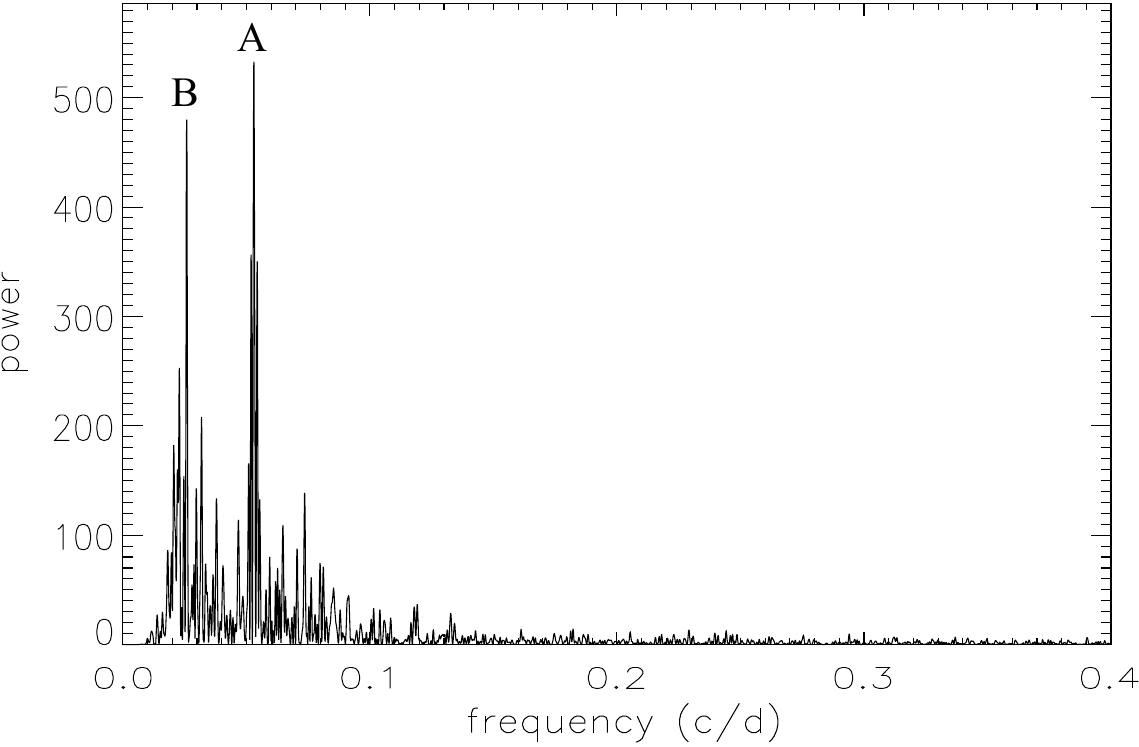}	
    \caption{{\it Upper panel}: light curve with additional treatment, wavelet local and global spectra of KIC 1151468, which was classified as periodic by \citet{Maoz2015}.  Contour levels are 90$\%$, 80$\%$, 70$\%$, ..., 20$\%$, and 10$\%$ of the map maximum. The variability period is illustrated by the red dashed line in the global spectra. A possible aliasing period is displayed in blue color. {\it Lower panel}: Lomb-Scargle periodogram for the same LC. The main peak depicted by the letter `A' corresponds to the period of 18.86 days, whereas letter `B' indicates the 39-days period.}
    \label{fig4}
\end{figure}

\begin{figure*}
	\includegraphics[width=\columnwidth]{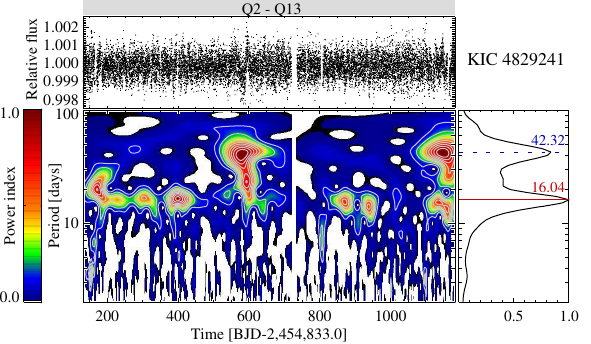}
	\includegraphics[height=4.75cm,width=\columnwidth]{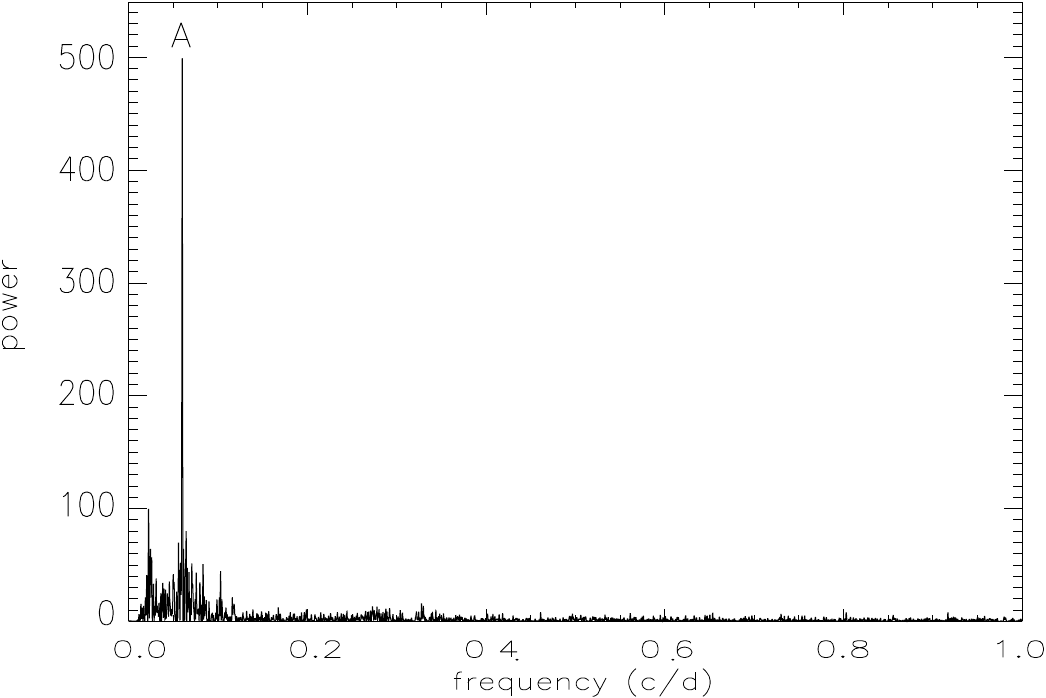}
	\includegraphics[width=\columnwidth]{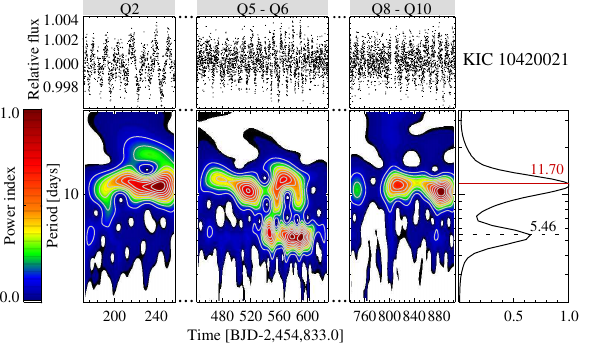}
	\includegraphics[height=4.75cm,width=\columnwidth]{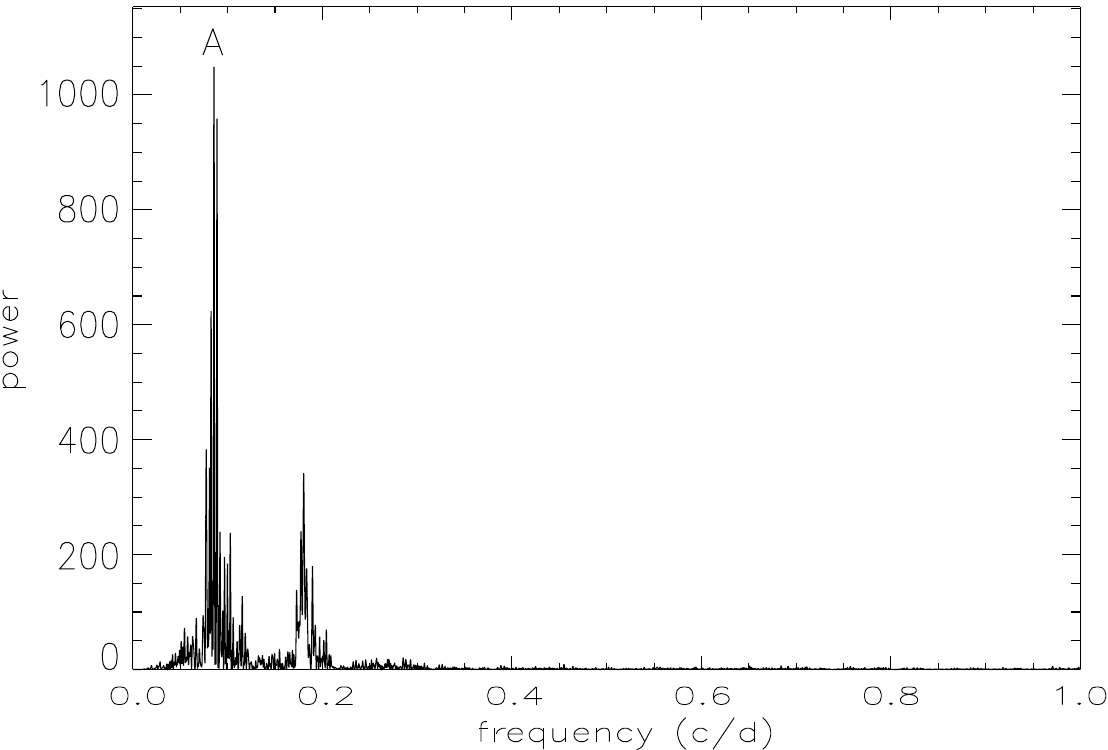}
	\includegraphics[width=\columnwidth]{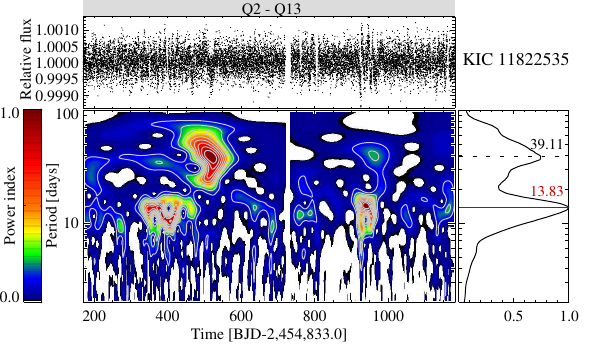}
	\includegraphics[height=4.75cm,width=\columnwidth]{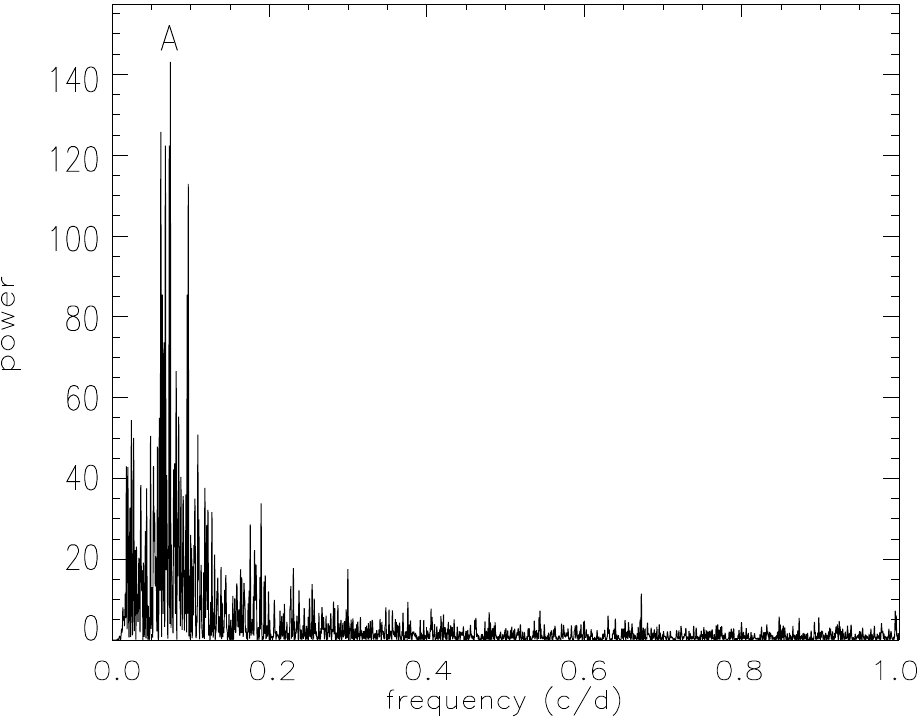}
    \caption{{\it Left panels}: light curves with additional treatment, wavelet local and global spectra of {\em Kepler} WD (from top to bottom) KIC 4829241, KIC 10420021, and KIC 11822535. Contour levels are 90$\%$, 80$\%$, 70$\%$, ..., 20$\%$, and 10$\%$ of the map maximum. The variability periods are illustrated by the red dashed lines in the global spectra. Possible aliasing periods are displayed in blue color. {\it Right panels}: respective Lomb-Scargle periodogram for the same WD. The main peak depicted by the A letter, corresponds to the main period obtained in the wavelet spectrum.}
    \label{fig5}
\end{figure*}

\subsection{KIC 11514682}

For this star the wavelet analysis reveals two periodicities, in contrast with the 9.89\thinspace days found by \citet{Maoz2015}. The wavelet map displayed in the upper panel of Figure \ref{fig4} shows a predominant period of 40.42\thinspace days and a secondary one at 17.59\thinspace days. These two periods are confirmed in the Lomb-Scargle periodogram displayed in the lower panel of Figure \ref{fig4}. The periods of 17.59 days and 40.42 days have approximately a 1:2 correspondence, respectively. This suggests that the first one is more likely a physical period and the latter one is a possible alias.

\subsection{KIC 4829241, KIC 10420021 and KIC 11822535}

The stars KIC 4829241, KIC 10420021 and KIC 11822535 were classified as non-periodic by \citet{Maoz2015}. Instead, our study suggests a periodic or quasi-periodic modulation well describing a variability signature in the wavelet representations. We attribute to this variability periodicities ranging between 11 and 16\thinspace days, which is reasonable for WD \citep[e.g.][]{Kawaler2015,Valeev2017,Braker2017}. In fact, these periodicities are more persistent in KIC 4829241 and KIC 10420021 than in KIC 11822535, even though also noticeable in the latter star. We do not expect to find magnetic cool spots signatures for KIC 4829241 and KIC 11822535 considering their high $T_{eff}$ and their log\thinspace g values \citep{2015ApJ...812...19T}. However, a transient signature for KIC 10420021 is conceivable in view of its $T_{eff}$ and log\thinspace g being close to the limit value \citep{2015ApJ...812...19T} to still envisage the presence of a convection zone at the WD surface. 

Combining the surface gravities and the effective temperatures, derived spectroscopically by \citet{2018MNRAS.476..933H}, with evolutionary grids \citep{1999MNRAS.303...30B}, we are able to estimate the mass of these three WDs. KIC 4829241 is a WD of mass $\sim$ $0.45 - 0.80$ $\mathrm{M_{\odot}}$. The Lomb-Scargle periodogram for this WD shows a period of 16.68\thinspace days with an amplitude of 286 ppm, as indicated by the letter A in the upper-right panel of Figure \ref{fig5}. Following \citet{Maoz2015}, given its mass, rotation combined with an accretion of gas from the interstellar medium (ISM) and the material carried on the magnetic poles generating hotspots, with an accretion luminosity of  $\Delta L_{acc} \sim 1.4\times 10^{29}$ erg s$^{-1}$, could explain the periodic modulation. Figure \ref{fig5} (upper-left panel) displays the wavelet analysis, from which we noticed a long-lasting periodicity of 16.04\thinspace days over the time observation. With a lower amplitude, we also found a 40-day period which could be considered as a possible aliasing period. The 16-day period characterizes the WD photometric modulation since it is the main period obtained in both Fourier- and wavelet-based techniques. Based on a rough FAP estimation (see Sect. 2), these periods likely lie above the confidence level. 

Left side middle of Figure \ref{fig5} displays the time series and wavelet local and global spectra of KIC 10420021. We can clearly track a long-lasting periodicity of 11.70\thinspace days and a second of 5.46\thinspace days. Although, the latter is only observed with high amplitude in quarters Q5 - Q7. Its time evolution could reveal a likely emergence of cool magnetic spots in the WD surface. Their short lifetime is still viable considering the WD effective temperature of about $\mathrm{12\thinspace 000}$\thinspace K \citep{1995PASP..107.1047B} for which the weak magnetic field still inhibits convective motions. Both periodicities are in accordance with the values obtained through Fourier techniques. The main peak of a period of 11.68\thinspace days in the Lomb-Scargle periodogram is indicated by the letter A in the right-side middle of Figure \ref{fig5}. From the wavelet analysis, we consider the 12-day period related to the variability modulation, probably due to the WD rotation. Given its small mass of $\sim$ $0.45 - 0.60$ $\mathrm{M_{\odot}}$ estimated using the $T_{eff}$ calculated by \citet{2018MNRAS.476..933H}; KIC 10420021 could be part of WDs for which photometric variability is due to rotation plus UV line absorption plus optical fluorescence. 

KIC 11822535 has a mass ranging approximately from 0.45 to 0.60\thinspace $\mathrm{M_{\odot}}$. The wavelet analysis, shown in the lower-left panel of Figure \ref{fig5}, is very similar to that of KIC 4829241. However, for this star, the main period of 13.83\thinspace days describes short quasi-periodic variations over the entire time span with lower amplitude. In accordance, a period of 13.56\thinspace days corresponds to the main peak of the respective Lomb-Scargle periodogram (Figure \ref{fig5}, lower-right panel) with a small amplitude of 74.3 ppm. As the effective temperature of this WD is higher than 30 kK, the quasi-periodic variability observed in the wavelet map could not be associated with magnetic spots, but it is possible to have a photometric modulation due to the rotation. This modulation could result from WD rotation plus magnetic dichroism or UV line absorption plus optical fluorescence.

\section{Conclusions}\label{conclusions}

We have carried out a wavelet analysis of the variability behavior of 14 stars observed by Kepler mission, previously analyzed by \citet{Maoz2015} using Fast Fourier Transform approach. We confirm the variability for all the 7 stars classified as periodic by those authors. Nevertheless, for one WD, KIC 11514682, we have found different periodicities. Let us underline that the variable star KIC 6862653 early classified as a WD, turned out to be a carbon-rich helium-dominated subdwarf \citet{2018MNRAS.476..933H}. We also identified noticeable variability for 3 WD stars reported by \citet{Maoz2015} as non-periodic. The stars KIC 4829241, KIC 10420021 and KIC 11822535, present wavelet maps revealing clear signature of periodicities, ranging approximately from 11 to 16\thinspace days. These are then placed within the list of WD with periodic modulations, while for KIC 3427482, KIC 7129927, KIC 9139775 and KIC 10198116 we confirm their non-periodic behavior.  

The 10 stars here studied present different features in their time-frequency representations. For KIC 5769827 and KIC 116047812 a particular semi-regular colour-filled contour, lasting all along their time span, leads to assume the rotation modulation as the most probable agent of such local map signature. As the KIC 116047812 effective temperature is about 9 000 K, it is very likely that the observed rotation modulation is combined with cool magnetic spots. Otherwise, the maps of KIC 6669882, KIC 6862653 and KIC 11337598 reveal a short-term variability which is hardly unhidden only by visual inspection of the data. The wavelet procedure helps on the identification of such short periodicities, which define a variability modulation characterized by their persistence, despite their low amplitude, along the time span. The 5-day periodicity found for the KIC 8682822 WD is considered to be less reliable because it only appears during almost 20\thinspace days of the total time span of 360\thinspace days. It was necessary to cut the entire light curve in small time intervals and apply the wavelet tratment in each dataset to pick up short timescales. 

The time-frequency behaviors for KIC 11514682, 4829241 and 11822535 are very similar. The main peaks in the wavelet global spectra ranging from 13 to 18 days describe a semi-regular track probably associated to the WD variability. Other periods higher than 11\thinspace days are in those cases considered as aliasing of the most predominant period. Finally, KIC 10420021 classified as a DA-type WD by \citet{Ostensen2011} presents expected conditions for cool magnetic spots, in view of its effective temperature value close to the threshold limit value for which their atmospheres are expected to be fully radiative. This is noticeable in its wavelet analysis, for which we found two main periodicities of 11.70 and 5.46\thinspace days.  

As underlined, the wavelet procedure offers a unique possibility for LC analysis because we can map short- and long-term features from the energy distribution of each LC computed as a local wavelet spectrum. In the present study, in addition to an enlarged treatment of the LCs, performed to avoid possible distortions in the spectral wavelet maps and changes in the power spectra, our wavelet approach was able to reveal new variability traces in the analyzed stellar sample.

\section*{Acknowledgements}

Research activities of the observational astronomy board at the Federal University of Rio Grande do Norte are supported by continuous grants from the brazilian funding agencies CNPq, and FAPERN and by the INCT-INEspa\c{c}o. SRdeL acknowledges CAPES graduate fellowships. JPB and ADC acknowledge a CAPES/PNPD fellowship. ICL acknowledges a postdoctoral fellowship from the Brazilian agency CNPq (Science Without Borders program, Grant No. 207393/2014-1). DBdeF acknowledges financial support from the Brazilian agency CNPq-PQ2 (grant No. 306007/2015-0). We also thank the reviewer, Prof. S. O. Kepler, for very useful comments to the original manuscript. This paper includes data collected by the Kepler mission. Funding for the Kepler mission is provided by the NASA Science Mission Directorate. The data presented in this paper were obtained from the Mikulski Archive for Space Telescopes (MAST). STScI is operated by the Association of Universities for Research in Astronomy, Inc., under NASA contract NAS5-26555.










\label{lastpage}
\end{document}